\documentstyle[prl,aps,epsfig]{revtex}

\def\Dsl{\,\raise.15ex\hbox{$/$}\mkern-13.5mu D}

\def\beq{\begin{equation}}
\def\eeq{\end{equation}}
\def\bea{\begin{eqnarray}}
\def\eea{\end{eqnarray}}

\newcommand{\vsl}{\slash \kern-12pt\hbox{\it v}}
\newcommand{\sla} [1]{\slash \kern-0.25cm\hbox{\it #1}}

\def\arto{ {\,\,\lower .8ex\hbox {$\longrightarrow
 \atop a \rightarrow 0$}\,\,}}

\def\R1{\varepsilon_1}
\def\E8{\varepsilon_8}

\newcommand{\bd}{\begin{displaymath}}
\newcommand{\ed}{\end{displaymath}}

\newcommand{\be}{\begin{equation}}
\newcommand{\ee}{\end{equation}}
\newcommand{\bi}{\begin{itemize}}
\newcommand{\ei}{\end{itemize}}

\def\npb#1#2#3{Nucl.\ Phys.\ {\bf B#1} (#2) #3}
\def\plb#1#2#3{Phys.\ Lett.\ {\bf B#1} (#2) #3}

\def\prl#1#2#3{Phys.\ Rev.\ Lett.\ #1 (#2) #3}

\begin{document}
\vspace*{0.5cm}

\setcounter{page}{1}

\begin{center}
\bf{\huge 
 A transparent expression  of the $A^2$-Condensate's renormalisation}
\end{center}  
\vskip 0.8cm
\begin{center}{\bf  Ph. Boucaud$^a$, F. De Soto$^{b}$, A. Le Yaouanc$^a$, J.P. Leroy$^a$, 
J. Micheli$^a$, H. Moutarde$^c$,
O. P\`ene$^a$, \\ 
J. Rodr\'{\i}guez--Quintero$^d$   
}\\
\vskip 0.5cm 
$^{a}$ {\sl Laboratoire de Physique Th\'eorique~\footnote{Unit\'e Mixte 
de Recherche du CNRS - UMR 8627}\\
Universit\'e de Paris XI, B\^atiment 210, 91405 Orsay Cedex,
France}\\
$^b${\sl Dpto. de F\'{\i}sica At\'omica, Molecular y Nuclear \\
Universidad de Sevilla, Apdo. 1065, 41080 Sevilla, Spain} \\
$^c$ Centre de Physique Th\'eorique Ecole Polytechnique, 
91128 Palaiseau Cedex, France \\
$^d${\sl Dpto. de F\'{\i}sica Aplicada e Ingenier\'{\i}a el\'ectrica \\
E.P.S. La R\'abida, Universidad de Huelva, 21819 Palos de la fra., Spain} \\
\end{center}

\medskip

\begin{abstract}
We give a more transparent understanding of the vacuum expectation
value of the
renormalised  local operator $A^2$ by relating it to the gluon propagator
integrated over the momentum. The quadratically divergent  perturbative
contribution is subtracted and the remainder, dominantly due to the ${\cal
O}(1/p^2)$ correction to the perturbative propagator at large $p^2$ is
logarithmically divergent.  This provides a transparent derivation of
the fact that this ${\cal O}(1/p^2)$ term is related to the vacuum
expectation value of the  local $A^2$ operator and confirms
a previous claim based on the  operator product expansion (OPE) of the gluon
propagator. At leading logarithms the agreement is quantitative, with a standard
running factor, between the local $A^2$ condensate renormalised  as described
above and the one renormalised in the OPE context. This result
supports the claim that the BRST invariant Landau-gauge $A^2$ condensate
might play an important role in describing the QCD vacuum.

\noindent P.A.C.S.: 12.38.Aw; 12.38.Gc; 12.38.Cy; 11.15.H
\end{abstract}

\section{Introduction}

In a series of lattice
studies~\cite{Boucaud:2000ey,Boucaud:2000nd,Boucaud:2001st,Boucaud:2002nc} 
the gluon propagator in QCD has been computed at large momenta, and it was 
shown that its behavior was compatible with the perturbative expectation
provided a rather large $1/p^2$ correction was considered. In an OPE approach 
this correction
has been shown~\cite{Boucaud:2000nd,Boucaud:2001st}
 to stem from an $A^2$  gluon condensate which does not vanish
 since the calculations are performed in the Landau gauge. It was also 
 claimed~\cite{Boucaud:2002nc} that this condensate might be related
 to instantons. 
 
The role of such a condensate in the non-perturbative properties of QCD, in
particular its relation to confinement, has been studied by several
groups~\cite{Kondo:2001nq,Zakh}. Of course any physics discussion about
the $A^2$ condensate  necessitates a clear definition of what we speak about, 
i.e. it needs a well  defined renormalisation procedure to define the
renormalised  local $A^2$ operator, since $A(0)^2$ is a quadratically
divergent quantity as can easily be seen in perturbation theory. 
A renormalisation of $A^2$ was defined 
in~\cite{Boucaud:2000nd,Boucaud:2001st} within the OPE context which we
now briefly summarise. It uses the notion of ``normal order
product'' in a ``perturbative vacuum'' which is 
annihilated by the fields $A$.  It implies that
  $<:A(0)^2:>_{\rm pert}=0$ in the perturbative vacuum\footnote{The 
symbol ``:~...~:'' represents
  the normal ordered product in this perturbative vacuum.}. The 
  contribution to $<:A(0)^2:>$ in the true QCD vacuum
   is then of non-perturbative origin.
  It has only logarithmic divergences and
  it is multiplicatively renormalised. Of course this notion
  of a perturbative vacuum in which Fock expansion could be performed
  has not a very transparent physical meaning especially in a non-perturbative 
  context such as the numerical Euclidean path integral method.

$A^2$ is not a gauge invariant operator but the bare $A^2$ condensate
is a very special object since, by definition, it is a minimum of the 
gauge orbit~\cite{Zakh}. In other words, some important physics seems 
to lie beneath the BRST invariance of $A^2$ in Landau gauge.
The authors of ref.~\cite{Kondo:2001nq} discussed on the
generalised\footnote{Landau gauge is recovered in the limit $\alpha
  \to 0$} composite operator $A_\mu A^\mu + 2 i \alpha \bar{c} c$,
which is BRST invariant in the manifestly Lorentz covariant gauge, and
examined the survival of this invariance after renormalisation.
In this note, although in a different context, we also
examine the same point: the subtle relationship between the minimum of
bare $A^2$ in the gauge orbit and any gauge-independant physical
phenomenology associated to the renormalised condensate\cite{Zakh}, 
emerging for instance from the OPE 
analysis~\cite{Boucaud:2000ey,Boucaud:2000nd,Boucaud:2001st,Boucaud:2002nc}. 
To this aim, we will derive the
renormalised  $A^2$ vacuum expectation value without using  the normal ordering
but using only  the OPE expansion of the gluon propagator \footnote{Of course the
normal ordering has been used  in~\cite{Boucaud:2000nd,Boucaud:2001st} to compute
the anomalous dimension of $A^2$ and the Wilson coefficient $c_2$}. It will
provide a more transparent definition, related directly to a quantity which is
actually measured. 

We start from the observation that the non renormalised $<A(0)^2>$ is related to
the integral of the gluon propagator over momentum. Hence it is expected that the
non-perturbative contribution to $A^2$ has to do with the non-perturbative
contribution to the gluon propagator. The latter contains precisely $1/p^2$ 
contributions due to the $A^2$ condensate at large momenta, and also strong
deviations from perturbative QCD at small momenta,  see
fig. \ref{fig} (taken from
\cite{Boucaud:2000ey,Boucaud:2000nd,Boucaud:2001st,Boucaud:2002nc}).  
How does this fit together?

\section{Bare, perturbative and non-perturbative $A^2$}

\begin{figure}[hbt]
\begin{center}
\mbox{\epsfig{file=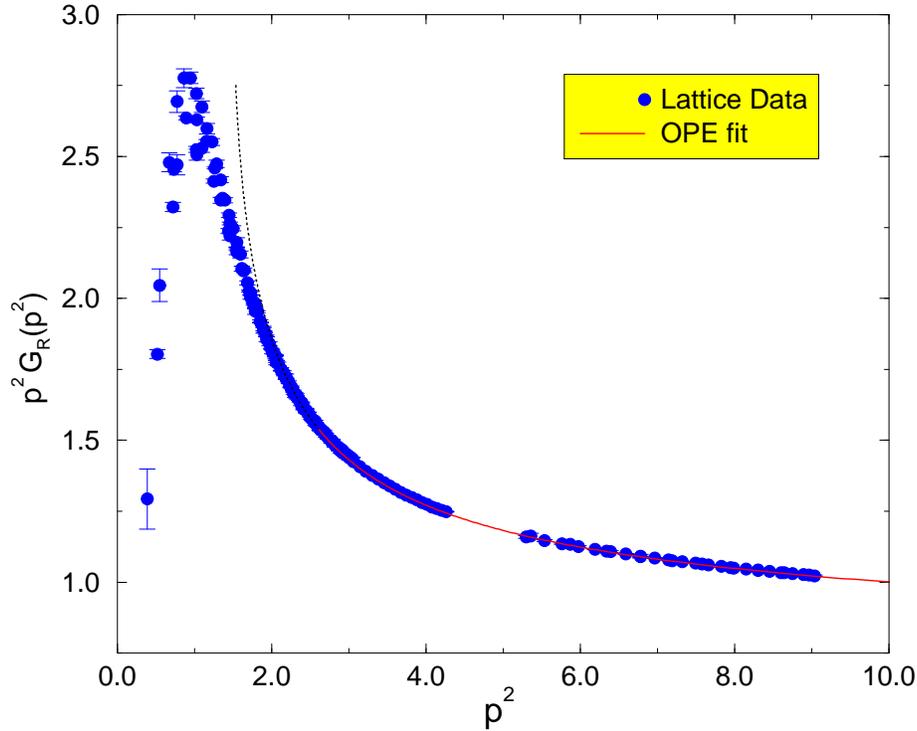,height=10cm}}
\caption{\small Gluon propagator extracted from lattice calculations
  renormalised at $\mu=10$ GeV and plotted between 0 and 9 GeV. The
  curve corresponds to the 
  fit written in eq. (\ref{gren}). It results that the infrared
  cut-off $p_{\rm min}$ can be safely taken around 2.6 - 3.0 GeV.} 
\label{fig}
\end{center}
\end{figure}

It is possible in principle from lattice calculations to define
the non-perturbative gluon propagator in the Landau gauge.
Lattice calculations provide the bare gluon propagator.
From the gluon propagator computed with a series of different 
values of the lattice spacing one can in principle 
compute the renormalised gluon propagator from zero momentum
up to as large a momentum as one wishes. An example of such a
non-perturbative propagator is shown in fig. \ref{fig}.
We can choose for example the MOM renormalisation scheme\footnote{Notice that the chosen
renormalisation scheme is not relevant in our argument in this
paper, but we clearly need a scheme in which non-perturbative quantities 
coming from lattice simulations can be accommodated. MOM is one of the simplest.
On the contrary the $\overline {\rm MS}$ scheme does not fulfill this condition.}, 
such that

\bea\label{mom}
G_{\rm R}^{(2)}(p^2=\mu^2)=\frac 1 {\mu^2}  \ .
\eea
This implies a renormalisation of the gluon fields
\bea\label{z3}
A_{\nu\,{\rm R}} = Z_3(\mu)^{-\frac 1 2}A_{\nu\,{\rm bare}},\quad Z_3(\mu) \equiv
  \mu^2\,G_{\rm bare}^{(2)}(\mu^2)
\eea
The renormalisation constant $Z_3$  has to be understood as related 
to any  regularisation method and any value of the UV regulator
provided that the latter is larger than the momenta carried by
the gluons. The coupling constant is also renormalised in the MOM scheme.
The Yang-Mills theory is thus fully renormalised and from now on we
 will consider only renormalised  gauge fields and propagators.

The propagator is defined in Euclidean space
by
\bea
\int d^4 x e^{ip\cdot x} < A_{\mu\,{\rm R}}^a(0) A_{\nu\,{\rm R}}^b(x)> 
= \delta_{a,b} 
\left[ \delta_{\mu,\nu}- \frac {p_\mu p_\nu} {p^2}\right] G_{{\rm R}}(p^2)
\eea
Inverting the Fourier transform,
\beq\label{four}
\sum_{a,\mu}<A_{\mu\,{\rm R}}^a(0) A_a^{\mu\,{\rm R}}(0)> 
= \frac {3(N_c^2-1)}{(2\pi)^4}\int d^4 p G_{{\rm R}}^{(2)}(p^2) 
= \frac {3(N_c^2-1)}{16 \pi^2}\int p^2dp^2 G_{{\rm R}}^{(2)}(p^2)
\eeq

\vskip 1cm

This integral is quadratically divergent in the ultraviolet. 
Indeed, if the gauge fields and the coupling constant have been
renormalised, the local $A^2$ operator has not yet. Let us 
introduce an ultraviolet cut-off $\Lambda$ and define
\bea\label{master}
<\left(A_{\rm R}(\mu)\right)^2>_{\Lambda} = 
\frac{3(N_c^2-1)}{16\pi^2}\int_0^{\Lambda^2} p^2 dp^2 G_{\rm R}^{(2)}(p^2).
\eea
where $\left(A_{\rm R}(\mu)\right)^2$ refers to the square of the gauge fields
renormalised at the scale $\mu$,  but where $A^2$ has {\it not been renormalised
as a local product of operators}. The symbol "$<...>$" represents the vacuum
expectation value.  $\left(A_{{\rm R}}\right)^2$ is clearly an UV divergent
quantity. The index $\Lambda$ refers to the ultraviolet cut-off and $\mu$ to the
renormalisation point for the gauge fields and the coupling constant.  The
cut-off $\Lambda$ has nothing to do with the lattice cut-off $a^{-1}$. The
renormalisation in eqs.~(\ref{mom}) and (\ref{z3}) has eliminated any  dependence
in the different lattice spacings which have been used to produce the 
renormalised propagator. $\Lambda$ is introduced simply to control the quadratic
and logarithmic divergences we encounter here.

The dominant contribution to this integral is the perturbative one.
To separate the perturbative contribution from the non perturbative
we will now use the results of~\cite{Boucaud:2001st}
\beq\label{old}
p^2 G^{(2)}_{\rm R}(p^2,\mu^2)  
 \equiv \frac{p^2 G^{(2)}(p^2)}{\mu^2 G^{(2)}(\mu^2)}=
{c_0\left(\frac{p^2}{\mu^2},
\alpha(\mu)\right)}  + 
{c_2\left(\frac{p^2}{\mu^2},\alpha(\mu) \right)}  
\frac{< \left(A^2\right)_{\rm R}(\mu) >}{4(N_c^2-1)} \ \frac{1}{p^2} \; .
 \eeq
 where $G^{(2)}(p^2)$ is the bare propagator. This expansion 
 does not exactly separate the perturbative form the non-perturbative
 contribution because of the denominator $\mu^2 G^{(2)}(\mu^2)$ which contains
 a non perturbative contribution.  
It is therefore convenient to introduce a slightly different renormalisation
$R'$:  
\beq\label{gren}
p^2 G^{(2)}_{\rm R'}(p^2,\mu^2)  \equiv  \frac {p^2 G^{(2)}(p^2)}
{\mu^2 G^{(2)}_{\rm pert}(\mu^2)} =
\frac{c_0\left(\frac{p^2}{\mu^2},
\alpha(\mu)\right)} {c_0\left(1,
\alpha(\mu)\right)}\ + 
\frac {c_2\left(\frac{p^2}{\mu^2},\alpha(\mu) \right)} {c_0\left(1,
\alpha(\mu)\right)}\ 
\frac{< \left(A^2\right)_{\rm R}(\mu) >}{4(N_c^2-1)} \ \frac{1}{p^2} \; .
 \eeq
where $\left(A^2\right)_{\rm R}(\mu)$ represents the $A^2$ operator
renormalised as a local operator at the scale $\mu$.
 Here the denominator is only the
perturbative contribution to the Green function whence the first term 
in eq. (\ref{gren}) is purely perturbative: it runs perturbatively with
a perturbative MOM renormalisation condition at $p^2=\mu^2$. 
Let us define for simplicity the constant 
\beq\label{z0}
z_0 \equiv\frac 1 {c_0\left(1,
\alpha(\mu)\right)}=\frac{G^{(2)}(\mu^2)}{G_{\rm pert}^{(2)}(\mu^2)} = 1 + 
{\cal }O\left(\frac 1{\mu^2}\right).
\eeq 
From~\cite{Boucaud:2001st} the first term in eq. (\ref{gren}),  $z_0
\,c_0\left({p^2}/{\mu^2}, \alpha(\mu)\right)$, represents the three loop 
perturbative contribution and the second, $z_0\, c_2\left({p^2}/{\mu^2},
\alpha(\mu)\right)$, the leading logarithm Wilson coefficient of the ${\cal
O}(1/p^2)$ nonperturbative correction attributed to the vacuum expectation value
of the renormalised  local operator $< \left(A^2\right)_{\rm R}(\mu) >$.
 
 Let us now introduce some notations:
\beq
\left(A^2\right)_{\rm R}(\mu)=Z_{A^2}^{-1}(\mu) :A_{\rm bare}^2: = 
\widehat{Z}^{-1}(\mu) 
:\left(A_{\rm R}(\mu)\right)^2: \ , 
\eeq

\noindent where $\widehat{Z}(\mu)\equiv Z_3^{-1}(\mu) Z_{A^2}(\mu)$, and 
\bea
\gamma_{A^2}\left(\alpha(\mu)\right)&\equiv&
\frac{d}{d\ln{\mu^2}} \ln{Z_{A^2}(\mu)}
=-{35 N_C \over 12}\frac{\alpha(\mu)}{4 \pi} +... \ ,  \nonumber \\
\widehat{\gamma}\left(\alpha(\mu)\right) \
&\equiv&\frac{d}{d\ln{\mu^2}} \ln{\widehat{Z}(\mu)} = 
\ - \widehat{\gamma}_0  \ \frac{\alpha(\mu)}{4 \pi} +...\ =
\ - \frac{3 N_C}{4} \ \frac{\alpha(\mu)}{4 \pi} +... \ \ 
\eea
where the symbol $:...:$ represents the normal ordered product in
the perturbative vacuum\footnote{The $:~...~:$
symbols have been erroneously omitted in~\cite{Boucaud:2001st}.
}. 
Our main goal in this paper is to understand better
the connection between $< \left(A^2\right)_{\rm R}(\mu) >$ 
defined in ~\cite{Boucaud:2001st} and the 
$\left(A_{\rm R}(\mu)\right)^2$ object considered here. 

The expansion in (\ref{gren}) is only valid
above some momentum $p \ge p_{\rm min}$. Typically we have taken 
$p_{\rm min} = 2.6$ GeV for our fits reported in 
\cite{Boucaud:2000ey}~-~\cite{Boucaud:2002nc}.

From eq. (\ref{master}), (\ref{gren}) we decompose 
\bea\label{decomp}
 <\left(A_{\rm R}(\mu)\right)^2>_{\Lambda} = 
 <\left(A_{\rm R}(\mu)\right)^2>^{{\rm pert}}_{\Lambda} +
  <\left(A_{\rm R}(\mu)\right)^2>^{{\rm OPE}}_{\Lambda} + 
 <\left(A_{\rm R}(\mu)\right)^2>^{{\rm IR}}_{\Lambda}
\eea
where
\bea\label{defpert}
<\left(A_{\rm R}(\mu)\right)^2>^{{\rm pert}}_{\Lambda} &=& 
\frac{3(N_c^2-1)z_0}{16\pi^2}\int_{p^2_{\rm min}}^{\Lambda^2} 
dp^2 c_0\left(\frac{p^2}{\mu^2},
\alpha(\mu)\right);\\ 
<\left(A_{\rm R}(\mu)\right)^2>^{{\rm OPE}}_{\Lambda} &=& 
\frac{3(N_c^2-1)z_0}{16\pi^2}\int_{p^2_{\rm min}}^{\Lambda^2} 
\frac {dp^2}{p^2} c_2\left(\frac{p^2}{\mu^2},
\alpha(\mu)\right)
\frac{< \left(A^2\right)_{\rm R}(\mu) >}{4(N_c^2-1)}; \nonumber \\ 
\label{defOPE}\\ \label{defIR}
<\left(A_{\rm R}(\mu)\right)^2>^{{\rm IR}}_{\Lambda} &=& 
\frac{3(N_c^2-1)}{16\pi^2}\int_0^{p^2_{\rm min}} p^2 dp^2 
G_{\rm R'}^{(2)}(p^2).
\eea
A few  comments are in order here. 
 $<\left(A_{\rm R}(\mu)\right)^2>^{{\rm pert}}_{\Lambda}$
corresponds indeed  to the perturbative computation of the vacuum expectation
value of $A^2$, i.e. to the connected diagrams with no external legs and with
one $A^2$ inserted.  However, the coupling constant and the gluon fields in
the diagrams  have been consistently renormalised at the scale $\mu$.  To leading
order eq. (\ref{defpert}) leads to
\bea\label{c0}
<\left(A_{\rm R}(\mu)\right)^2>^{{\rm pert}}_{\Lambda}
{\kern 2.em \hbox{$\to$} \kern -2.0em\raise -1.5ex  \hbox{$ \Lambda \to\infty$} }
\frac{3(N_c^2-1)}{16\pi^2} \ \Lambda^2 \nonumber \\
\times \left\{ \left(\frac{\ln \left(\frac{\Lambda}{\Lambda_{\rm QCD}}\right)}
{\ln \left(\frac{\mu}{\Lambda_{\rm QCD}}\right)}
\right)^{\frac {\gamma_0}{\beta_0}} \left( 1+ O\left( \frac{1}{\ln\left(
\frac{\Lambda}{\Lambda_{\rm QCD}}\right)}\right)\right) + 
O\left( \frac{p_{\rm min}^2}{\Lambda^2} \right)
\right\}
\eea
which diverges more than quadratically.
Note that the dependence in $p^2_{\rm min}$ is subdominant.

 In equation (\ref{defOPE}) the left hand side has been defined from the
 decomposition  of the integral (\ref{master}) according to  (\ref{gren}).
The right hand side contains  $< \left(A^2\right)_{\rm R}(\mu) >$ 
already discussed.
 The latter
is just a number  which factorises out of the integral in   (\ref{defOPE}).
We thus see that $<\left(A_{\rm R}(\mu)\right)^2>^{{\rm OPE}}_{\Lambda}$  and 
$< \left(A^2\right)_{\rm R}(\mu) >$
 are proportional. 

Our next task is to compute the proportionality coefficient and to compare 
$<\left(A_{\rm R}(\mu)\right)^2>^{{\rm OPE}}_{\Lambda}$ 
 with the other subleading term, 
  $<\left(A_{\rm R}(\mu)\right)^2>^{{\rm IR}}_{\Lambda}$. 
From~\cite{Boucaud:2001st} and  eq. (\ref{z0}) we know that 
$z_0= 1 + {\cal O}(1/\mu^2)$. Our calculation of the integral in eq. (\ref{defOPE})
being performed to leading logarithm we will take $z_0 c_2= c_2$ in the 
 following. From (23) in~\cite{Boucaud:2001st},
\bea
c_2\left(\frac{p^2}{\mu^2},\alpha(\mu)\right)  =  12 \pi \alpha(p) \ 
\left( \frac{\alpha(p)}{\alpha(\mu)} 
\right)^{- \frac{\widehat{\gamma}_0} {\beta_0}}.
\eea
Let us also recall
\bea
Z_3(\mu) \propto  \left(\alpha(\mu)\right)^{\frac {\gamma_0}{\beta_0}},\quad 
 < \left(A^2\right)_{\rm R}(\mu) > \propto 
 \left(\alpha(\mu)\right)^{-\frac {\gamma_{A^2}}{\beta_0}}
\eea
with 
\bea\label{gamma0}
\beta_0=11,\quad\gamma_0 = 13/2,\quad \gamma_{A^2}= \frac {35}{4},\quad
\widehat{\gamma}_0 = \gamma_{A^2} - \gamma_0 = \frac {9} {4}
\eea

From eq. (\ref{defOPE}) and the leading logarithm relation
\beq
d p^2/p^2= d \log(p^2) \simeq -\frac {4 \pi}{\beta_0} \frac {d \alpha}{\alpha^2}
\eeq
we find 
\bea\label{int}
<\left(A_{\rm R}(\mu)\right)^2>^{{\rm OPE}}_{\Lambda}& =& \frac{3(N_c^2-1)}{16\pi^2}\,\frac{(12 \pi)}
{(\alpha(\mu))^{- \frac{\widehat{\gamma}_0} {\beta_0}}} 
\,\frac{< \left(A^2\right)_{\rm R}(\mu) >}{4(N_c^2-1)}\, 
\frac {4 \pi}{\beta_0}\, \int^{\alpha(p_{\rm min})}_{\alpha(\Lambda)} 
d\alpha \alpha^{-1 - \frac{\widehat{\gamma}_0} {\beta_0}} \nonumber 
\\ &=& 
< \left(A^2\right)_{\rm R}(\mu) >
\left[ \left(\frac{\alpha(\Lambda)}{\alpha(\mu))}\right)
^{- \frac{\widehat{\gamma}_0} {\beta_0}}-
\left(\frac{\alpha(p_{\rm min})}{\alpha(\mu))}\right)
^{- \frac{\widehat{\gamma}_0} {\beta_0}}\right]
\eea 
It is interesting to notice that the coefficient $\beta_0/\widehat{\gamma_0}$
 stemming from the integration over $\alpha$ is exactly compensated by 
 the prefactors outside  the integral, the origin of which does not
 appear at first sight to be related to the anomalous dimension of $A^2$.
 Had we taken any other anomalous dimension instead of $\widehat{\gamma_0}$,
 say some $\gamma'$, we would have ended with a constant $9/(4 \gamma')$ in
 front of the r.h.s of eq. (\ref{int}). 
 
In the large $\Lambda$ limit, $\alpha(p_{\rm min})\gg\alpha(\Lambda) $
whence, since $\widehat{\gamma}_0$ is positive, the main result of
this note comes from: 
\bea\label{final1}
<\left(A_{\rm R}(\mu)\right)^2>^{{\rm OPE}}_{\Lambda} 
{\kern 2.em \hbox{$\simeq$} \kern -2.0em\raise -1.5ex  \hbox{$ \Lambda \to\infty$} }
< \left(A^2\right)_{\rm R}(\mu) >
 \left(\frac{\alpha(\Lambda)}{\alpha(\mu))}\right)
^{- \frac{\widehat{\gamma}_0} {\beta_0}} \ .
\eea 
To leading logarithms and keeping $\mu$ fixed, 
\bea
<\left(A_{\rm R}(\mu)\right)^2>^{{\rm OPE}}_{\Lambda}\propto \, 
\alpha(\Lambda)^{-\frac{\widehat{\gamma}_0} {\beta_0}} 
{\kern 2.em \hbox{$\to$} \kern -2.0em\raise -1.5ex  \hbox{$ \Lambda \to\infty$} }
 \infty \ \ .
\eea
On the other hand, form eq. (\ref{defIR}) 
\bea\label{IRzero}
   <\left(A_{\rm R}(\mu)\right)^2>^{{\rm IR}}_{\Lambda} = {\rm constant}.
\eea
since it does not depend on $\Lambda$. It results that 
$<\left(A_{\rm R}(\mu)\right)^2>^{{\rm OPE}}_{\Lambda}$
is dominant over $<\left(A_{\rm R}(\mu)\right)^2>^{{\rm IR}}_{\Lambda}$
 in the decomposition (\ref{decomp}). This dominance will lead to the
 coming interpretation (next section) of the decomposition
 (\ref{decomp}), which is indeed the main result of this note.
Of course, we perform our analysis in the Landau
gauge because of its conceptual~\cite{Zakh} and numerical (Lattice
Green functions~\cite{Boucaud:2001st}) particular interest.
However the survival after renormalisation of BRST invariance in
covariant gauges for the generalised composite operator showed in 
ref.~\cite{Kondo:2001nq} seems to point out that an analogous
analysis, with similar results, for these gauges could be performed.  

As an interesting special case, if $\mu=\Lambda$
\bea\label{final2}
<\left(A_{\rm R}(\Lambda)\right)^2>^{{\rm OPE}}_{\Lambda}
{\kern 2.em \hbox{$\to$} \kern -2.0em\raise -1.5ex  \hbox{$ \Lambda \to\infty$} }
\,\,<\left( A^2\right)_{\rm R}(\Lambda)> \propto 
\left(\alpha(\Lambda)\right)^{-\frac {\gamma_{A^2}}{\beta_0}}
\eea

\section{Conclusion and discussion}

 Our conclusion is summarised in 
\bea\label{final}
< \left(A^2\right)_{\rm R}(\mu) >&\simeq& 
<\left(A_{\rm R}(\mu)\right)^2>^{{\rm OPE}}_{\Lambda} 
\left(\frac{\alpha(\Lambda)}
 {\alpha(\mu))}\right)
^{\frac{\widehat{\gamma}_0} {\beta_0}}
\nonumber \\ &\simeq&
 \left[ <\left(A_{\rm R}(\mu)\right)^2>_\Lambda
 - <\left(A_{\rm R}(\mu)\right)^2>^{{\rm pert}}_{\Lambda}\right]
 \left(\frac{\alpha(\Lambda)}
 {\alpha(\mu))}\right)
^{\frac{\widehat{\gamma}_0} {\beta_0}}. 
\eea
The notations being not conventional let us recall that the $<...>_\Lambda$'s in
the r.h.s represent the  gluon  propagator integrated over momentum up to an UV cut-off, 
$\Lambda$, see eqs. (\ref{master}) and (\ref{defpert}).   The gluon fields and coupling
constants are renormalised in all the terms appearing in these equation. Thus we
learn that the further renormalisation of the local operator $A^2$ proceeds by
substracting to the plain vacuum expectation value of $A^2$ the same object
computed in perturbation. This logarithmically divergent difference is then
renormalised by the powers of $\alpha$ in the r.h.s of eq.~(\ref{final}). Not
unexpectedly, we retrieve in essence the initial expression of  the
renormalisation of the $A^2$ operator through normal ordering (i.e. subtraction
of the perturbative v.e.v), followed by the multiplicative, logarithmic
renormalisation $Z_{A^2}$. But apart from a non trivial consistency check,
involving in particular the detailed expression of the  Wilson coefficient, we
obtain an expression which is more transparent, since it only involves a measurable
quantity, the integral over the renormalised propagator.

Eq. (\ref{final}) presents a separation between  perturbative and non-perturbative
contributions to the integrated propagator i.e. to $A^2$. Of course, such a 
separation depends on the renormalisation scheme, and on the order in perturbation
theory in which the Green functions are computed. It is also well known
that summing to infinity the perturbative series may generate renormalons
which behave like non-perturbative condensates. To avoid any such problem
we stick to a finite order in the perturbative series.  Furthermore, if the 
quantitative separation between  perturbative and non-perturbative
contributions depends on these prescription, the results summarised in eq
(\ref{final}) {\it do not depend on them} provided that we use {\it the same}
 scheme and order when computing both sides of eq.~(\ref{final}). Of course the
 anomalous dimensions to leading logarithms do not either depend on them.

This simple result has several interesting consequences.  First, it has been
advocated~\cite{{Boucaud:2002nc}} that the $A^2$ condensate could be dominantly due to the
contribution to the path integral of semi-classical gauge field configurations
such as instantons liquids. It is useful to consider this hypothesis through a
background field picture, i.e. factorising the path integral into an integral
over semi-classical gauge field configurations, and for each value of these an
integral over quantum fluctuations around this background configuration.  
It means that the hermitian matrix $A_\mu$ is decomposed into:
\bea\label{back}
A_\mu = B_\mu + Q_\mu(B)\quad A^2 = B^2 + \{B.Q\}_+ + Q^2(B)
\eea
$B_\mu$ being the background, assumed to be non-perturbative, and $Q_\mu$
the quantum fluctuations assumed to be perturbative. $\{B.Q\}_+ \equiv B.Q + Q.B$.
In principle, $Q_\mu$ depends on $B_\mu$ and differs from the quantum
fluctuations around the trivial vacuum $B_\mu=0$  which is what perturbative
QCD computes. The hypothesis that $<\left(A^2\right)_{\rm R}>$ is 
due~\footnote{This discussion is qualitative and 
we do not know how to define rigorously the corresponding scale $\mu$.
We therefore prefer to omit writing $\mu$ here.} to
 these semi-classical gauge configurations is translated into:
  $\left(A^2\right)_{\rm R}\simeq B^2$. From eqs.
 (\ref{final}), (\ref{back}) 
 \bea
 <\left(A^2\right)_{\rm R}>\simeq <B^2> \simeq <A^2> - <Q^2(B=0)>
 \eea 
 i.e. that $(Q^2(B)-Q^2(B=0))$ is subleading\footnote{
 If $B$ is a classical solution of the field equations, the term
 linear in $Q$ will vanish. $B$ should be close to such
 a solution and we therefore neglect $\{B.Q\}_+$.}. 
In other words  the 
 dependence of $Q_\mu$ on $B_\mu$ is subleading. The hard quantum fluctuations
 are not sensitive to the soft background field.
 
A most interesting consequence of our result is related to some
discussions in \cite{Zakh}. These authors
extend to QCD some remarks stemming from compact $U(1)$. They attribute
a special role to the $A^2$ condensate, even if a gauge dependent quantity, by
arguing that $A^2$ in the Landau gauge is the minimum of $A^2$ on the gauge
orbit. One difficulty in this argument is the following: Fixing the Landau gauge
amounts to minimize the $<A^2_{{\rm bare}}>$ while  the condensate refers
to some renormalised quantity free of the quadratic and logarithmic divergences.
 In compact $U(1)$ life is simpler:
\beq\label{zakh}
<A^2_{{\rm bare}}>= <A^2_{{\rm pert}}> + <A^2_{{\rm nonpert}}>,
\eeq
the perturbative theory is trivial and the nonperturbative contribution is due 
grossly speaking to
the topology. A phase transition when the coupling  constant
varies allows to measure directly the non perturbative contribution.  We refer to
\cite{Zakh} for more details. Our result eq. (\ref{final})
exhibits in QCD, up to subleading contributions,
 a linear decomposition similar to  eq. (\ref{zakh}), although such 
 a similarity is not at all obvious at first
sight. 
The next question could be whether in some sense the  $<A^2>^{{\rm OPE}}$
computed in the Landau gauge is the minimum of some quantity on the gauge orbit. 

 Last but not least, let us simply say that the result in eq. (\ref{final})
provides a fairly simple understanding of what the $A^2$ condensate is. {\it It
confirms that indeed the ${\cal O}(1/p^2)$ correction to perturbative  QCD at
large momenta has to do with the $A^2$ condensate.} Indeed, if one start with
some doubt about the relation of the r.h.s of eq. (\ref{defOPE}) with an $A^2$
condensate, just considering it as an unidentified $1/p^2$ contribution, we
end-up with the conclusion that it yields a non-perturbative contribution
to the  $A^2$ v.e.v. The fact that in our derivation
this term has precisely the anomalous dimension of an $A^2$ condensate
comes form the fact that  $c_2$ in the r.h.s of eq. (\ref{defOPE}) has been
computed under the assumption that it is due to an $A^2$ condensate, an
assumption which has been shown to fit fairly well the lattice data.
Had we used another scale dependence for $c_2$ we would have ended with
a wrong scale dependence for the resulting non-perturbative contribution
 to the  $A^2$ v.e.v. We would have also ended with a constant different
 from 1 in front of the r.h.s of eq. (\ref{final}), see the discussion 
 following eq. (\ref{int}). Thus the picture is fully consistent.

\section{Acknowledgments} 
 We are grateful to V.I. Zakharov for illuminating. This work was
 supported in part by the European Union Human Potential Program under
 contract HPRN-CT-2000-00145, Hadrons/Lattice QCD.

 discussions.


\vspace*{0.5cm}
\begin{thebibliography}{99}

\bibitem{Boucaud:2000ey}
P.~Boucaud {\it et al.},
JHEP {\bf 0004}, 006 (2000)
[arXiv:hep-ph/0003020];
D.~Becirevic, P.~Boucaud, J.~P.~Leroy, J.~Micheli, O.~Pene, 
J.~Rodriguez-Quintero and C.~Roiesnel,
Phys.\ Rev.\ D {\bf 61}, 114508 (2000)
[arXiv:hep-ph/9910204];
Phys.\ Rev.\ D {\bf 59}, 094509 (1999) 
[arXiv:hep-ph/9903364].

\bibitem{Boucaud:2000nd}
P.~Boucaud, A.~Le Yaouanc, J.~P.~Leroy, J.~Micheli, O.~Pene and J.~Rodriguez-Quintero,
Phys.\ Lett.\ B {\bf 493}, 315 (2000)
[arXiv:hep-ph/0008043];
F.~De Soto and J.~Rodriguez-Quintero,
Phys.\ Rev.\ D {\bf 64}, 114003 (2001)
[arXiv:hep-ph/0105063].


\bibitem{Boucaud:2001st}
P.~Boucaud, A.~Le Yaouanc, J.~P.~Leroy, J.~Micheli, O.~Pene and J.~Rodriguez-Quintero,
Phys.\ Rev.\ D {\bf 63}, 114003 (2001)
[arXiv:hep-ph/0101302].

\bibitem{Boucaud:2002nc}
P.~Boucaud {\it et al.}, Phys. \ Rev. \ D {\bf 66}, 034504
[arXiv:hep-ph/0203119];
[arXiv:hep-ph/0205187].

\bibitem{Kondo:2001nq}
K.~I.~Kondo,
Phys.\ Lett.\ B {\bf 514}, 335 (2001)
[arXiv:hep-th/0105299];
K.~I.~Kondo, T.~Murakami, T.~Shinohara and T.~Imai,
Phys.\ Rev.\ D {\bf 65}, 085034 (2002)
[arXiv:hep-th/0111256].

\bibitem{Zakh} 
        F.V. Gubarev, V.I. Zakharov, \plb{501}{2001}{28}, [arXiv:hep-ph/0010096]; 
        F.V.Gubarev, L. Stodolsky, V.I. Zakharov, \prl{86}{2001}{2220}, [arXiv:hep-ph/0010057].
        F.V. Gubarev, M.I. Polikarpov, V.I. Zakharov, [arXiv:hep-ph/9908292];
        K.G. Chetyrkin, S. Narison, V.I. Zakharov, \npb{550}{1999}{353}, [arXiv:hep-ph/9811275];



\end{thebibliography}
\end{document}